\documentclass[journal=jpcbfk,manuscript=article]{achemso}

\usepackage[version=3]{mhchem} 
\usepackage{braket}

\author{Zahra Shamsi}
\affiliation[University of Illinois at Urbana-Champaign, Urbana, IL]
{Department of Chemical and Biomolecular Engineering, University of Illinois at Urbana-Champaign, Urbana, IL 61801, USA}
\author{Kevin J. Cheng}
\affiliation[University of Illinois at Urbana-Champaign, Urbana, IL 61801, USA]
{Center for Biophysics and Quantitative Biology, University of Illinois at Urbana-Champaign, Urbana, IL 61801, USA}

\author{Diwakar Shukla}
\email{diwakar@illinois.edu}
\affiliation[University of Illinois at Urbana-Champaign, Urbana, IL 61801, USA]
{Center for Biophysics and Quantitative Biology, University of Illinois at Urbana-Champaign, Urbana, IL }
\alsoaffiliation[University of Illinois at Urbana-Champaign, Urbana, IL]
{Department of Chemical and Biomolecular Engineering, University of Illinois at Urbana-Champaign, Urbana, IL}
\alsoaffiliation[University of Illinois at Urbana-Champaign, Urbana, IL]
{Department of Plant Biology, University of Illinois at Urbana-Champaign, Urbana, IL}

\title[Reinforcement Learning]{REinforcement learning based Adaptive samPling: REAPing Rewards by Exploring Protein Conformational Landscapes}

\abbreviations{}

\begin{document}

\begin{abstract}
One of the key limitations of Molecular Dynamics simulations is the computational intractability of sampling protein conformational landscapes associated with either large system size or long timescales. To overcome this bottleneck, we present the REinforcement learning based Adaptive samPling (REAP) algorithm that aims to efficiently sample conformational space by learning the relative importance of each reaction coordinate as it samples the landscape. To achieve this, the algorithm uses concepts from the field of reinforcement learning, a subset of machine learning, which rewards sampling along important degrees of freedom and disregards others that do not facilitate exploration or exploitation. We demonstrate the effectiveness of REAP by comparing the sampling to long continuous MD simulations and least-counts adaptive sampling on two model landscapes (L-shaped and circular), and realistic systems such as alanine dipeptide and Src kinase. In all four systems, the REAP algorithm consistently demonstrates its ability to explore conformational space faster than the other two methods when comparing the expected values of the landscape discovered for a given amount of time. The key advantage of REAP is on-the-fly estimation of the importance of collective variables, which makes it particularly useful for systems with limited structural information.
\end{abstract}

\section{Introduction}

\par\noindent Molecular dynamics (MD) simulations have rapidly advanced into an invaluable tool for understanding the structure-function relationship in biological molecules.\cite{dror2012,adcoc2006, shukla2015markov,lane2012milliseconds} Although they aid our understanding of intricate biomolecular dynamics, the bottleneck lies in the amount of computational resources available to the researcher. In common practice, running MD simulations on non-specialized computing hardware allows for nanoseconds worth of data per day.\cite{dror2012}  The reality is that the salient protein conformational changes can occur at millisecond and even longer timescales; a six or greater order of magnitude difference in terms of ns\cite{adcoc2006}, which can cost up to years worth of simulation time. Examples include the transport cycle for membrane transporter proteins\cite{jiang2011large,moradi2013mechanistic}, protein folding\cite{lindorff2011fast, voelz2010molecular,lane2012milliseconds,lapidus2014complex}, and large-scale conformational changes involved in cell signaling \cite{lemmo2010, dror2012,kohlhoff2014cloud,lawrenz2015cloud,shukla2015chapter,shukla2016conformational,vanatta2015network, moffett2017molecular}. A number of enhanced sampling methods have emerged to address this computational drawback of conventional simulations. Two general classes exist among these methods; one class requires the specification of reaction coordinate,  i.e. a function of system degrees of freedom that guides the simulation to reach a desired end state by enhancing sampling along the reaction coordinate. This class can further be broken down into two subclasses, either by biasing the underlying potential along the reaction coordinate (e.g. steered MD,\cite{grubm1996} metadynamics\cite{Huber1994}, Temperature Accelerated MD\cite{abrams2010large}, umbrella sampling\cite{kastner2011umbrella}) or perform unbiased adaptive sampling using the reaction coordinates as a metric.\cite{weber2011characterization, bowman2010enhanced, singhal2004using}. The second class of techniques encourages exploration of the conformational landscape in all directions by modifying the overall Hamiltonian (e.g. accelerated MD,\cite{hamel2004} replica exchange MD\cite{sugit1999}, or weighted-ensemble simulations\cite{Huber1996}). Depending on the scientific goal, the usefulness of each class of techniques will differ. Several techniques exist that combine the ideas from these methods to achieve enhanced sampling efficiency. For example, Preto and Clementi introduced a new method called Extended DM-d-MD\cite{Preto2014}, that enhances the sampling of MD trajectories within areas that are typically difficult to sample such as the barriers between metastable regions. It accomplishes by iteratively restarting simulations so as to obtain a uniform distribution along the first two diffusion coordinates. Since the diffusion coordinates are obtained from post-processed simulation data, \textit{a priori} reaction coordinate information is not needed to perform the method. Similarly, the iMapD\cite{Chiavazzo2016} method attempts to efficiently explore the free energy surface of a system using an adaptive exploration strategy; it iteratively starts new simulations at the boundary points of a lower dimensional space (in their case, diffusion coordinates), and outwardly explores the space until new metastable configurations are detected. Another method, named SGOOP\cite{Tiwary2016}, attempts to find the best linear combination of a pre-selected set of reaction coordinates using maximum path entropy estimates. This newly generated coordinate can then be used to sample along using one of the enhanced sampling method mentioned above.

\par For this paper, we will focus on the issue with the first class which requires a set reaction coordinates as an input. Essentially, this class of methods can only succeed by knowing which reaction coordinates are relevant for sampling \textit{a priori}. In the context of biomolecular simulations, reaction coordinates are observables that capture the progress of proteins undergoing conformational change between different states. For instance, the distance between two residues can serve as reaction coordinates such that they only approach each other when the protein is in an activated” state. These computational techniques work by sampling the conformational space preferentially along the reaction coordinate, ``pushing” the protein process towards some final state of interest. 

\begin{figure}[th]
\centering
\includegraphics[width=\linewidth]{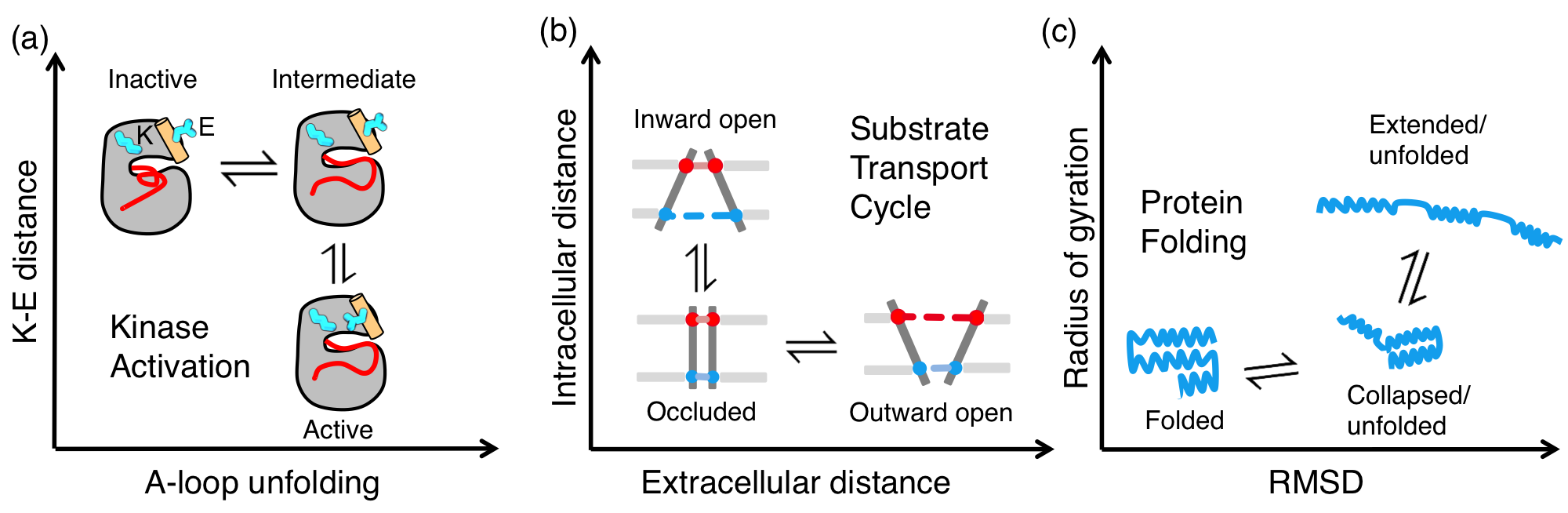}
\caption{The above figures illustrate that three significant biological processes can be projected onto an L-shaped landscape given the selection of appropriate reaction coordinates. These include: (a) The activation process in kinases, (b) transport cycle for transporter proteins, and (c) protein folding. In (a), the inactive kinase state is denoted as state 1, the intermediate state as 2, and the activated state as 3.}
\label{fig:sys}
\end{figure}

Reaction coordinates have proven useful for researchers as they help reduce the high dimensionality of the system. Since protein dynamics involves highly complex processes, it is desirable to project this high dimensional space onto reaction coordinates that simplify the simulation data without losing essential information regarding conformational. Furthermore, it is not uncommon to characterize the conformational dynamics as a projection onto two reaction coordinates (i.e. on 2-dimensional space), nonetheless, one is not limited to a 2-D projection, but is obviously preferred as it makes interpretation easier.  This is especially useful if the protein under investigation undergoes a series of intermediate steps to achieve some final state. For instance, many protein kinases (proteins involved in signaling via phosphotransfer) only become active after two molecular events occur: the A-loop unfolding, and the formation of the K-E salt bridge after the $\alpha$c-helix rotates to form a K-E salt bridge\cite{shukla2014activation,moffett2017molecular,meng2013locking,meng2016transition}. In other words, while one reaction coordinate changes, the other remains relatively constant.  Plotting the progression of these events gives rise to an ``L-shaped" landscape. Kinases are not the only biological systems that can be projected onto an L-shaped landscape using two reaction coordinates, these include membrane transporter proteins\cite{Chen2015,wang2015neurotransmitter} and protein folding\cite{voelz2010molecular,dror2012,lane2012milliseconds,lapidus2014complex} (see Fig. \ref{fig:sys} to visualize landscape).

\par\noindent It is evident from these three landscapes that as one reaction coordinate is important for sampling, the other becomes less relevant. To illustrate this point concretely, consider the landscape associated with c-Src kinase activation.\cite{shukla2014activation,moffett2017molecular,meng2013locking,meng2016transition} If simulations started with a kinase in the inactive state (Fig. \ref{fig:sys} (a), top left), the optimal sampling strategy would be to first explore in the positive x direction (rightward) along the A-loop unfolding coordinate. Once state 2 has been reached, the x coordinate no longer becomes relevant for sampling (since the A-loop has completely unfolded) and now the y coordinate (the K-E bond distance) becomes the optimal direction to sample. By sampling along this L landscape, we can be assured that the most relevant protein conformations have been sampled, i.e. the complete kinase activation cycle. However,  the only reason the optimal directions are known beforehand is from closely examining differences between crystal structures captured in various states. If in this example the protein were sampled without well-established structural information, the only option is to use a “brute-force” approach by which the landscape must be explored in all directions when there is clearly an optimal path in this case. 

\begin{figure}
\begin{center}
\includegraphics[width=1\textwidth]{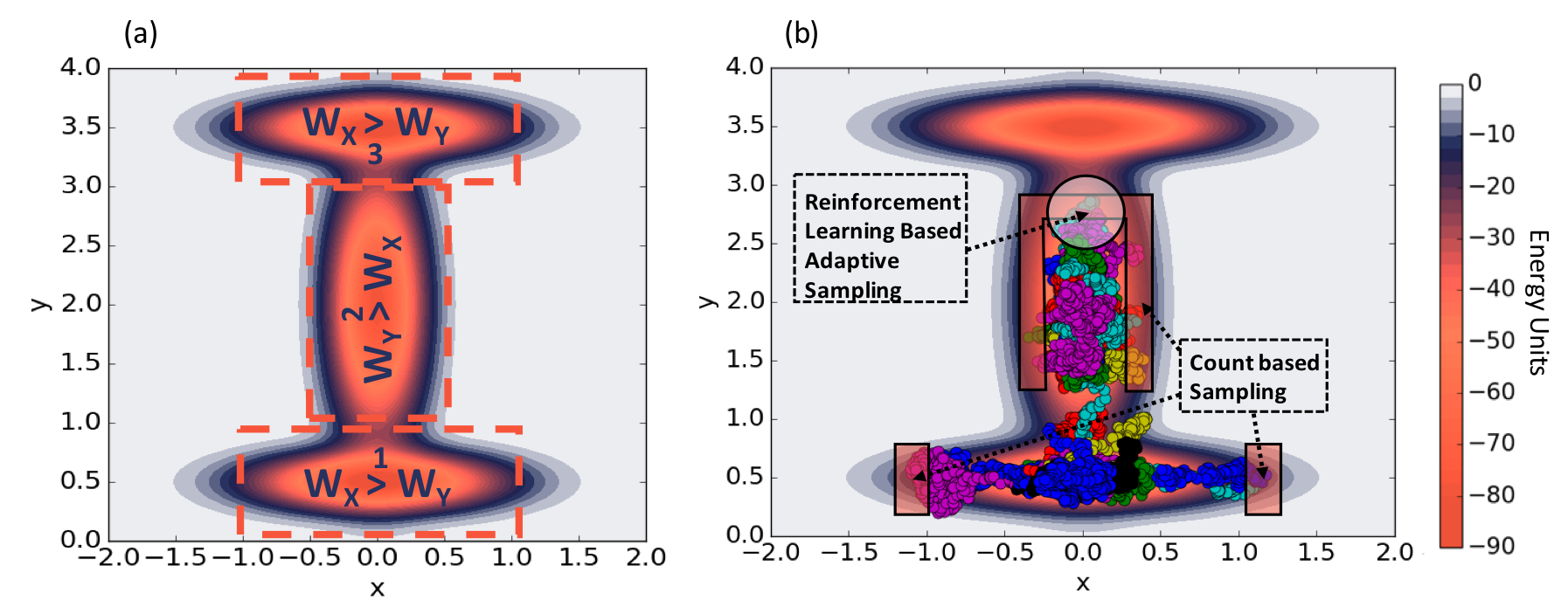}
\end{center}
\caption{{(a) The 'I' (Illinois) landscape illustrates that the local optimal sampling strategy changes depending which basin that is currently being sampled (each labeled 1, 2, and 3). The importance of each reaction coordinate is denoted as weights that are updated in each iteration of the REAP algorithm, $W_X$ and $W_Y$. (b) Given that sampling occurs in basin 2, the orange regions shows structures selected from count based sampling are not optimal for reaching basin 3. Instead, REAP is able to identify the appropriate structures (white circular highlight) that facilitate sampling along Y, eventually reaching basin 3. }}
\label{Aim1_Preliminary1}
\end{figure}

\par\noindent The concept of optimal sampling along some reaction coordinate at different points on the landscape has led us to develop the REAP (\textbf{RE}inforcement learning based \textbf{A}daptive sam\textbf{P}ling) algorithm that is ``smart" enough to determine the relative importance of each reaction coordinate as it explores conformational space.  Reinforcement learning (RL) constitutes a significant aspect of the artificial intelligence field with numerous applications ranging from finance to autonomous vehicles \cite{Wulfmeier2017}. It is based on the Pavlovian conditioning and control theory, where the feedback from the environment is learned to maximize the accumulated award. REAP takes principles from the field of RL \cite{szepesvari2010algorithms, gosavi2009reinforcement} by which an agent (or learning system) takes actions in an environment to maximize a reward function. In this study, the action is picking new structures to start a swarm of simulations, while the reward is a mathematical function proportional to how far reaction coordinates sample the landscape (see REAP Algorithm section for more details). The agent keeps track of which direction is most rewarding, allowing it choose the optimal sampling strategy. In other words, the agent attempts to find the path of least resistance. 

\par\noindent The REAP algorithm builds upon count-based adaptive sampling method. Both methods perform the following: 1) Run a series of short MD simulations from a collection of starting structures. 2) Cluster the proteins based on reaction coordinates of interest. 3) Pick structures from these clusters according to some sampling criterion to start new simulations. The difference between these methods resides in step 3, least count based adaptive sampling chooses new structures based on least populated states, while REAP chooses new structures based on a reward function. This reward function is dependent on weights (a parameter representing how important a reaction coordinate is) and how on the landscape the new simulation data samples compared to the current data. The advantage of REAP over adaptive sampling is that it decreases the chance of choosing structures for the next round of simulations that are irrelevant for sampling given the current state. Consider the \textbf{`I'} (Illinois) potential in Fig. \ref{Aim1_Preliminary1} (a) where the importance of the reaction coordinates X and Y changes in each basin.
In the situation, where sampling occurs in basin 2, a count-based \cite{weber2011characterization,singhal2004using} adaptive sampling would give equal importance to both X and Y, allowing structures in the orange regions to be chosen for the next round of sampling. The disadvantage is that a lot of these structures are irrelevant towards reaching the final area of sampling, basin 3. The REAP algorithm is able to identify that the most important structures, that is, the white highlighted region (Fig \ref{Aim1_Preliminary1} (b)) since they facilitate sampling along the Y direction. As a result, exploring low-energy, biologically relevant regions of the landscape becomes faster, effectively saving precious computational resources for the user.  

\par\noindent The use of reward functions to increase the efficiency of sampling has been implemented in other studies as well. For example, Zimmerman and Bowman \cite{zimmerman2015fast} have developed a goal-oriented sampling method to search conformational space for structures with desirable observables. The reward function in this case is maximized by taking the gradient of the structural metric of interest. Furthermore, Perez et al. \cite{Perez2017} have used the concept relevant to RL such as ``explore-and-exploit'' \cite{berry1997bandit} to enhance conformational exploration using data derived from experiments. REAP differs from both of these methods since it does not require modification of the original Hamiltonion or \textit{a priori} information regarding which physical properties should be maximized or minimized such as RMSD, residue pair distance, solvent accessible area, etc. The only input needed is a list of \textit{possible} reaction coordinates. 

\par\noindent This paper discusses and outlines the basic algorithm of REAP, then evaluates its performance compared to conventional single long trajectories (SL) and least counts sampling (LC) using two idealized potentials; an L and a circular landscape. It is then applied to alanine dipeptide MD simulations and Src kinase. The kinase system was sampled using a kinetic monte carlo sampling scheme based on markov state models obtained from a previous study. \cite{shukla2014activation} For each case, we plotted the distribution of landscape discovered using repeated simulation trials. The expected values for the REAP distributions were consistently higher than LC and SL, suggesting that REAP is a successful improvement of LC since it explores new areas of conformational space more efficiently. To avoid terminological confusion, we will interchangeably use reaction coordinate (RC) and order parameter (OP) for the remainder of this article.

\section{REAP Algorithm}
Here, we present each steps involved in the implementation of the REAP algorithm. We also introduce the RL concept of a \textit{policy} which defines the agent's way of behaving at a given time. In a mathematical sense, the policy $\pi$ is the mapping between ``states'' belonging to the environment and ``actions'' to achieve the agent's goal $ \pi : S \rightarrow A$. Put differently, the policy tells the agent how to behave at any point in time. The environment is defined as the landscape that is to be explored; with the state $S$ defined as the set of all discovered points on the landscape or simply the current data available. The action $A$ is defined as the agent choosing protein structures to run more simulations on. The user can provide different policies which differ in the RCs provided. By employing the sampling algorithm below, the user can evaluate which of these different policies ensures the most reward while sampling and then evaluate which RCs are relevant for sampling. To avoid any misunderstanding, the definition of ``states'' ($S$) here should not be confused with the common usage familiar to biophysicists to represent one particular protein configuration. 

\begin{enumerate}

\item{Identify some sampling policy $\pi_K$ and its corresponding set of RCs $K = \{\theta_1, \theta_2, \ldots, \theta_k\}$. These RCs could be based on known or likely RCs associated with the conformational transition under investigation. Each policy differs depending on the set of RCs and how new protein structures are chosen for each round of simulation. In our implementation for this work, the sampling policy involves choosing structures based on least populated clusters (performed at step 5, denoted as $C_p$) and the reward function of each cluster (Eq. \ref{eq:reward})

\item{Set the weight $w_i$ for each $\theta_i \in K$ where $w_i \in [0,1]$. The initialization of each $w_i$ signifies which RC $\theta_i$ is important for the first round of sampling. Of course, if no prior knowledge is available regarding the importance of each weight, each $w_i$ can be fixed to the constant value of ${1}/{k}$, where $k$ is the total number of RCs for the given policy $\pi_K$. Every iteration of this algorithm produces a new state $S$ (the set of all discovered points), and since $w_i$ is different for each $S$, we will introduce new notation for the weights $w_{i}^{S}$.}

\item{Run simulations to generate a series of initial structures. This can be obtained either from a single trajectory, or from running short simulations from multiple structures that can be obtained from homology modeling, crystal structures, biased MD methods etc.}

\item{Cluster the data $S$ into a set of $L$ clusters $C = \{c_{1}, c_{2}, \ldots, c_{L}\}$. For each cluster $c_j \in C$, identify all the structures that are closest to the cluster $c_j$. The user could also assign a representative structure to each cluster e.g the centroid of each cluster. The goal of this step is to reduce the data size by clumping together structures in RC space.  Nonetheless, the clustering method can be arbitrarily chosen during this step.}

\item{Identify the set of clusters $C_p \subset C$ which contain the top least number of data points. The cardinality (size) of $C_p$ is at the discretion of the user. As mentioned in step 1, the set $C_p$ can be obtained using a different criteria.}

\item{Given the set of $K$ RCs for policy $\pi_K$, calculate the reward for each $c_m \in C_p$. 

\begin{equation}\label{eq:reward}
r^{K}(c_m)=\sum\limits_{i=1}^{k} w_i^S  \frac{\mid (\theta_i(c_m)- \langle \theta_i(C) \rangle\mid}{\sigma_i(C)} 
\end{equation}

Where $w_i^S$ represents the weight or importance of each RC for a given set of discovered points $S_{RL}$, $\theta_i(c_m)$ is the RC calculated for the cluster $c_m$, $\langle \theta_i(C)\rangle$ is the arithmetic mean of $\theta_i$ for all $c_j \in C$, and $\sigma_i(C)$ represents the standard deviation of $\theta_i$ for all $c_j \in C$. Vertical bars indicate the absolute value being taken.}

\item{Calculate the cumulative reward.
\begin{equation}\label{eq:cumulativereward}
R(C_p) = \sum\limits_{m = 1}^{\mid C_p \mid}r^K(c_m)
\end{equation}
Where the sum is over each element in the set $C_p$, and $\mid C_p \mid $ is cardinality of $C_p$.}
\item{The next step is to maximize Eq. \ref{eq:cumulativereward} by tuning the parameter $w_i^S$. This can be achieved by choosing from a myriad of optimization algorithms already implemented. In our case, we took advantage of the SciPy python library\cite{scipy} and used the Sequential Least SQuares Programming (SLSQP)\cite{kraft1988software} to find the optimal weights that maximize the cumulative reward. The following conditions were enforced as a constraint: $\sum_i w_i = 1$ and $\mid w_{i}^{t-1} - w_{i}^{t} \mid \leq \delta$, $\forall i$, where  $ 0 < \delta < 1 $. $t$ represents the current round of sampling while $t-1$ represents the previous round. We found these constraints to make the algorithm more robust. Given the updated weights, step 6 is repeated to find the new rewards} 
}

\item{Choose the structures from the clusters that give the highest reward to start new simulations given the updated weights. The two additional parameters, structures and clusters chosen with the highest reward, is up to discretion of the user.}

\item{Repeat steps 3-9 until the user deems the sampling is sufficient enough.}

\end{enumerate}

\par \noindent The primary reason for using least count adaptive sampling as specified in step 4 is because it is widely considered as the most efficient strategy for exploration of free energy landscapes.\cite{weber2011characterization, bowman2010enhanced, singhal2004using} Even if the reward function cannot properly assign rewards to states to achieve optimal sampling, the least-count adaptive sampling protocol will still be used. The crucial step in this algorithm is step 8, which estimates the relative importance of these RC as the agent explores the landscape. If, for example, some $\theta_i$ is provided that gives poor information on conformational changes (i.e. changes little for each round of simulations), the weights will eventually drop to zero from the optimization step. This essentially informs the user that the RC is not important for understanding the conformational changes of the protein. In theory, one can provide different policies $\pi_K$ with the same reward function to determine which RCs are most relevant for sampling by simply looking how each $w_i$ changes over time.

\section{Applications to Model Potentials}

\subsection{L-Shaped Potential}
To demonstrate that the REAP algorithm outperforms other sampling strategies, we will first consider an idealized system by which the time-evolution of two RCs, X and Y, are governed by the overdamped Langevin equation:
\begin{equation}\label{eq:lang}
\dot{\textbf{r}}(t) = - 1/\gamma \nabla V(\textbf{r}(t)) +  \eta(t)\sqrt{2\beta^{-1}\gamma}
\end{equation}

Here, $\dot{\textbf{r}}(t)$ denote time derivatives of the position vector $\textbf{r}$, $\nabla$ is the gradient operator, $\gamma$ is the friction coefficient, $\beta= 1/k_BT$ where $k_B$ is the Boltzmann constant and $T$ is the temperature in Kelvin, $\eta(t)$ represents a random force that models the collisions of molecules in a fluid. The random force obeys a Gaussian distribution with a zero mean and satisfies the following autocorrelation condition $ \Braket{\eta_i(t) \eta_j (t^\prime)} = \delta_{ij} \delta(t-t^\prime)$. \cite{fogedby1994langevin} 

The goal is to study the performance of this sampling method for a model potential representing landscapes common in protein conformational change, such as ion transports, activation processes, and protein folding (see Fig. \ref{fig:sys} for examples). We will first assume that the potential is L-shaped with five metastable states, i.e. long-lived intermediate states between the initial and final state. 

\begin{figure}[ht]
\centering
\includegraphics[width=\linewidth]{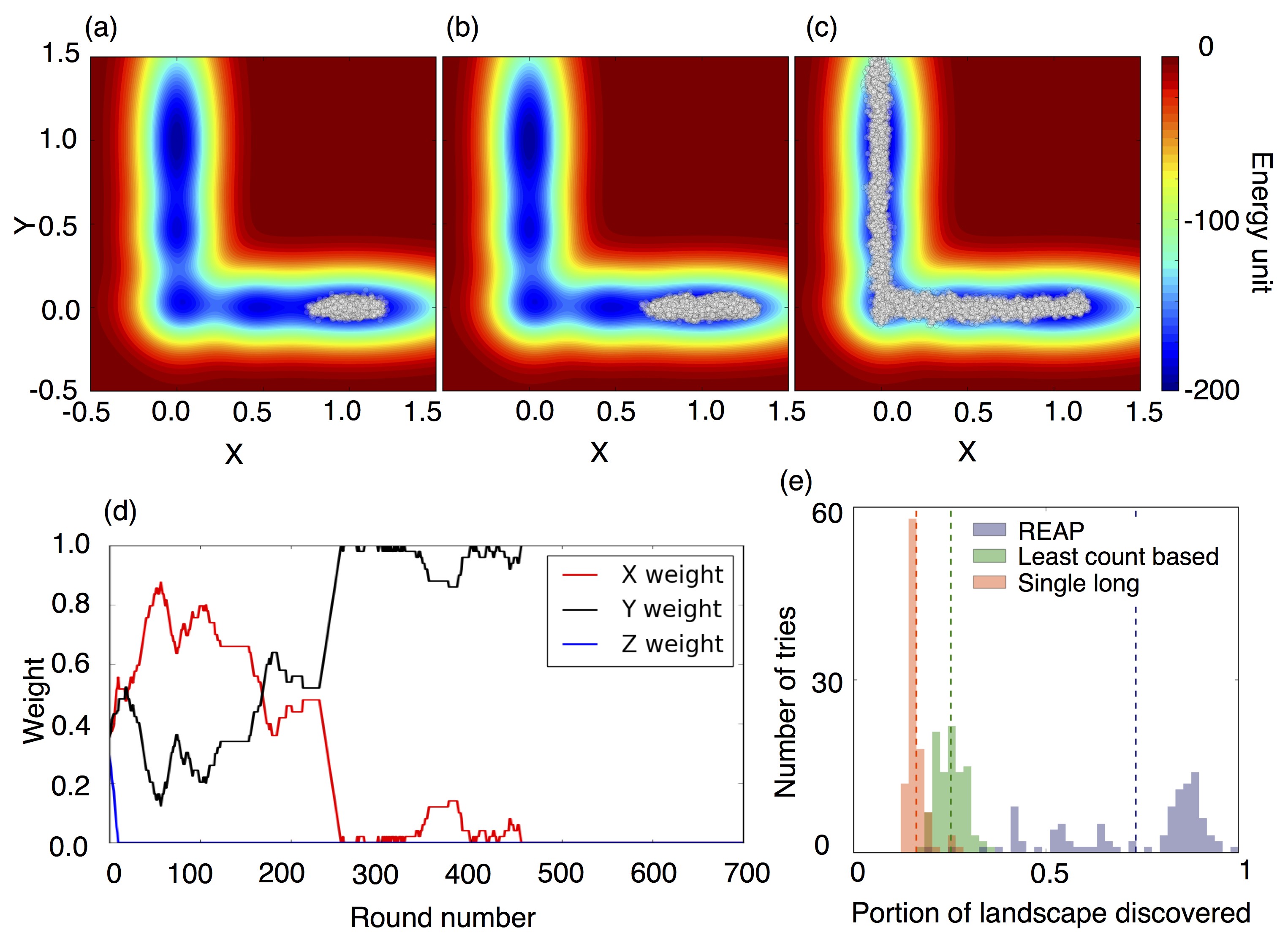}
\caption{Regions sampled using (a) single long trajectory, (b) least count based adaptive sampling, and (c) REAP algorithm methods performed on L-shaped potentials are shown with white circles on top of the potential. The white circles represents a data point generated from Eq. \ref{eq:lang} (d) Weights for each RC signify the importance of each RC depending on the round number (or iteration of the algorithm). The fluctuations of weights show that the algorithm is able to identify the importance of each weight. The weight of an additional RC orthogonal to X and Y, called Z, was expected and shown to go to zero. For more information on what the weights signify, see step 2 in the REAP Algorithm section. (e)  A plot showing the distribution for the portion of landscape discovered using REAP, LC, and SL sampling over 100 repeated trials. Dashed lines represent the expected value of each distribution.}
\label{fig:L}
\end{figure}

To compare each sampling method, we performed the exact same amount of simulation time on the L-shaped landscape for each sampling procedures; traditional simulation (single long trajectory or SL), least count based  adaptive sampling (LC), and the REAP algorithm \cite{weber2011} (Fig. \ref{fig:L} (a), (b), (c) respectively.).

All three simulations were initiated at the bottom right corner of the landscape (point (1.1 , 0) in Fig. \ref{fig:L}). These model simulations clearly demonstrate the advantage of using REAP for exploring the landscape. This is due to the algorithms ability to quickly identify the important directions of sampling. These directions are quantified as weights for each RC (see Eq. \ref{eq:reward}), facilitating immediate exploration to the left. This learning of directional importance can be visualized by looking at the change in weight values over time (see Fig. \ref{fig:L} (d)). We additionally included an unvarying OP, Z, to demonstrate the algorithms capability to identify its insignificance. At the very start of the simulations, REAP quickly assigns this additional OP with zero weight, thereby preventing further sampling along this coordinate. REAP's ability to identify unimportant OPs in a more realistic system, such as Src kinase, will be elaborated on in a later section.

In the first rounds of simulations, both X and Y directions are equally important (Fig. \ref{fig:L} (d) up to round ~10) as sampling in all directions gives equal rewards. It is not until trajectories reach the high energy states in the Y direction, where the X direction now becomes more rewarding than Y. As a result, the X weight increases while the Y weight decreases. When the trajectories reach the point where X no longer becomes rewarding near round 300 (point (0, 0) in Fig. \ref{fig:L} (c)), then the X weight decreases and Y weight dramatically increases until the trajectories reach the fifth basin (top left corner, point (0, 1.1) in Fig. \ref{fig:L} (c)).

To compare the performances between REAP, LC, and SL, we plotted the distribution of landscape discovered using 100 repeated simulation trials (Fig. \ref{fig:L} (e)). The distribution for SL and LC were concentrated around 0.2 and 0.25 respectively, while REAP is populated mostly at 0.8 with more distribution spread. Furthermore, the expected value for REAP is, at least, twice as much as LC and SL--suggesting that REAP is expected to perform better than the latter two. Additional plots showing the distribution for each individual OPs were calculated (SI Fig. 1 and 2), and the time evolution of average smallest value of Xs (SI Fig. 3). These different metrics provide a alternative perspective for showing how REAP outperforms the other two methods.

We further assessed how REAP performs with an additional, non-functional OP (i.e. an OP that does not change in time). Our results suggest that the distributions were very similar compared to using two OPs (SI Fig. 4). 

\subsection{Circular Potential}
\begin{figure}[ht]
\centering
\includegraphics[width=\linewidth]{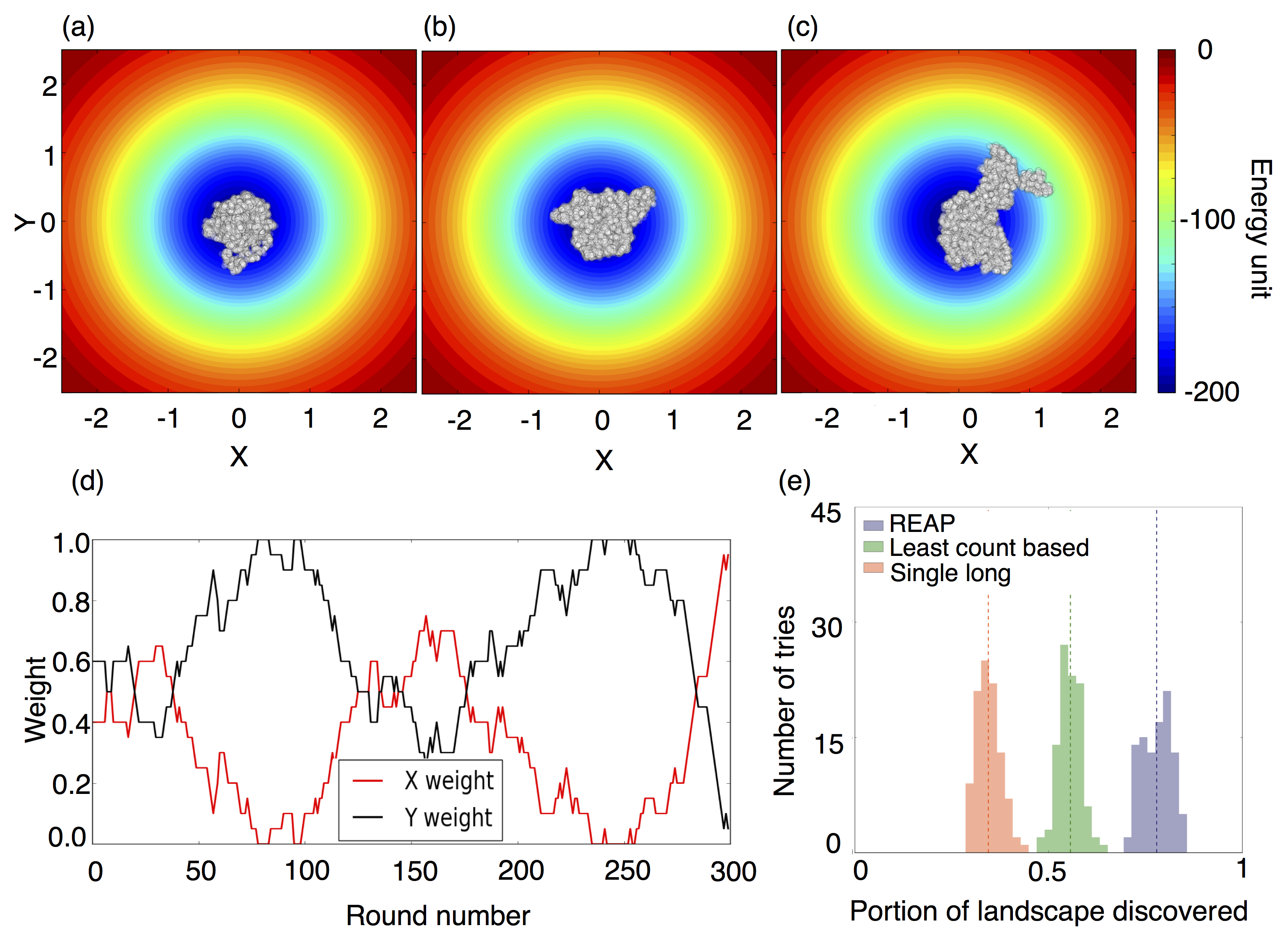}
\caption{The Regions sampled using (a) single long trajectory, (b) least count based adaptive sampling, and (c) REAP algorithm methods performed on circular potentials are shown with white circles on top of the potential. The white circles represents a data point generated from Eq. \ref{eq:lang} (d) Weights for each RC signify the importance of each RC depending on the round number (or iteration of the algorithm) (e) A plot showing the distribution for portion of landscape discovered for REAP, LC, and SL simulations using 100 repeated simulations. The expectation value of each distribution is depicted with dashed lines.}
\label{fig:c}
\end{figure}

In the second model system, we tested the performance of the algorithm in which there is no important direction of sampling at any given time. Thus, we considered a single circular potential with a single metastable state. The reason for choosing this model potential is to show that the REAP algorithm performance is the same as the least-count adaptive sampling for a system with no preferred direction of sampling. The dynamics were governed by Eq. \ref{eq:lang} with a single circular metastable state (Fig. \ref{fig:c}). 

We performed three sampling procedures: SL, LC, and REAP, with the same amount of sampling. The majority of generated points from traditional SL simulation and least count adaptive sampling were mostly confined to the  stable basin (Fig. \ref{fig:c} (a) and (b)) while the REAP algorithm explored more regions of the landscape (Fig. \ref{fig:c} (c)). The trajectories using REAP demonstrated some directionality, attempting to sample one direction at a time. The importance of each coordinate as a function of round number is illustrated in Fig. \ref{fig:c} (d). We again found that REAP outperforms both LC and SL for this landscape. This is evident from the distribution plots for portion of landscape discovered in Fig. \ref{fig:c} (e). Three distinct distributions arise from this landscape, with REAP's distribution shifted towards higher values than LC and SL. REAP again demonstrates better performance relative to the other sampling methods. Since model potentials are only insofar useful for conceptual demonstrations, we will further consider two molecular systems such as alanine dipeptide, a ubiquitous benchmark in the enhanced sampling literature, and Src kinase. 

\subsection{Alanine Dipeptide}

\begin{figure}[ht]
\centering
\includegraphics[width=\linewidth]{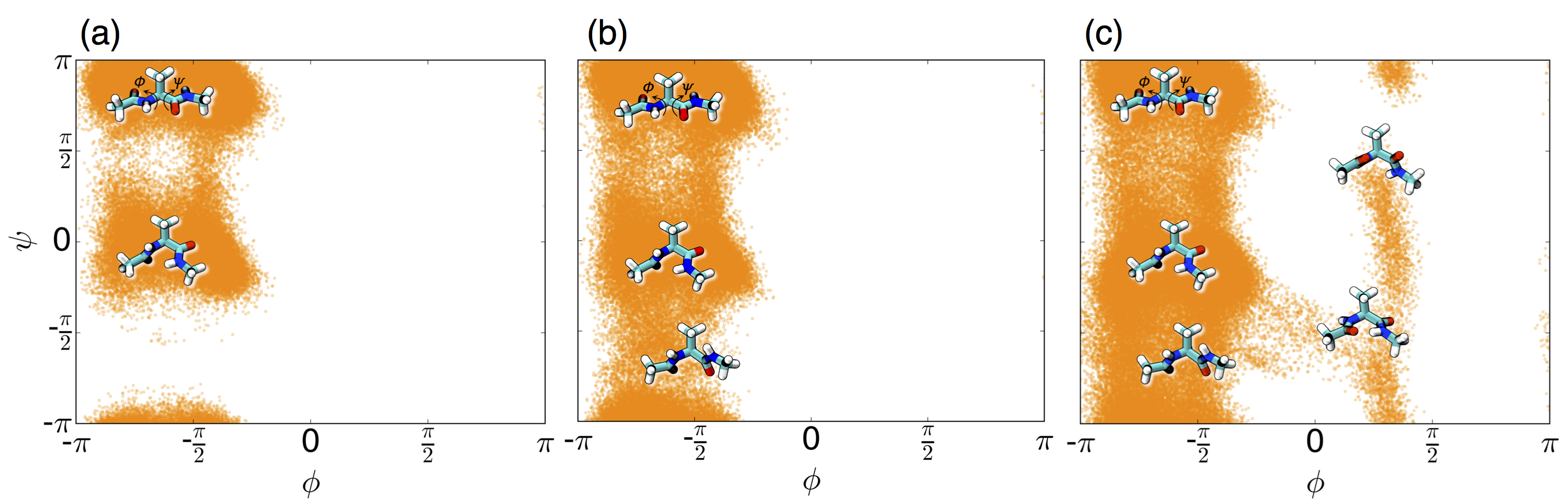}
\caption{Alanine dipeptide landscapes were generated using three simulation methods: (a) single long trajectory simulation, (b) least counts adaptive and (C) the REAP algorithm. Representative molecular structures are shown on top of the landscape to illustrate the new regions sampled using REAP within the same simulation time.}
\label{fig:ala}
\end{figure}

To illustrate that our algorithm remains effective when using MD to sample, we applied the REAP algorithm to alanine dipeptide. We performed a total of 2 ns simulations to sample the dihedral angle landscape using SL trajectories, LC, REAP (Fig. \ref{fig:ala}). With regards to simulation details: the starting structure of alanine dipeptide was obtained from the Python package MSMBuilder 3.8 \cite{harrigan2017msmbuilder}. The simulation was carried out using OpenMM.\cite{eastman2017openmm} The AMBER99SB forcefield \cite{hornak2006comparison} was used along with the TIP3P\cite{mark2001structure} water model. A cubic box with periodic boundaries was employed to model bulk solvent. The final system consisted of 1831 atoms and simulated at temperature of 300 K while using a Langevin integrator to propagate particle motion and regulate temperature. A friction coefficient of 1.0 / ps. A Monte Carlo barostat \cite{Chow1995} was used to maintain a pressure of 1 bar. To deal with long range interactions, a nonbonded cutoff of 10 $\AA$ was used with the Particle Mesh Ewald (PME) \cite{darden1993particle} method to calculate long distance interactions. The system was initially minimized for 1 ps (500 steps) then equilibrated for 200 ps. A total of 2 ns was generated from production MD runs. Trajectories were saved every 0.1 ps . A timestep of 2 fs and hydrogen bonds were constrained for the entire simulation.

Simulations started from the most stable state of the peptide at (-$\pi/2$, $\pi$) (Fig. \ref{fig:ala}). After 2 ns, SL of alanine dipeptide captured two metastable state in the landscape, but failed to sample new metastable states at ($\pi/4$,-$\pi/2$) revealed using REAP. LC improves on SL, as it is able to better sample the regions for $\psi = \pi/4$ and  $\psi = -\pi/2$. Additional landscapes are provided to visualize how the majority of REAP trials explored more regions (SI Fig. 5, 6, 7). 

\begin{figure}[ht]
\centering
\includegraphics[width=\linewidth]{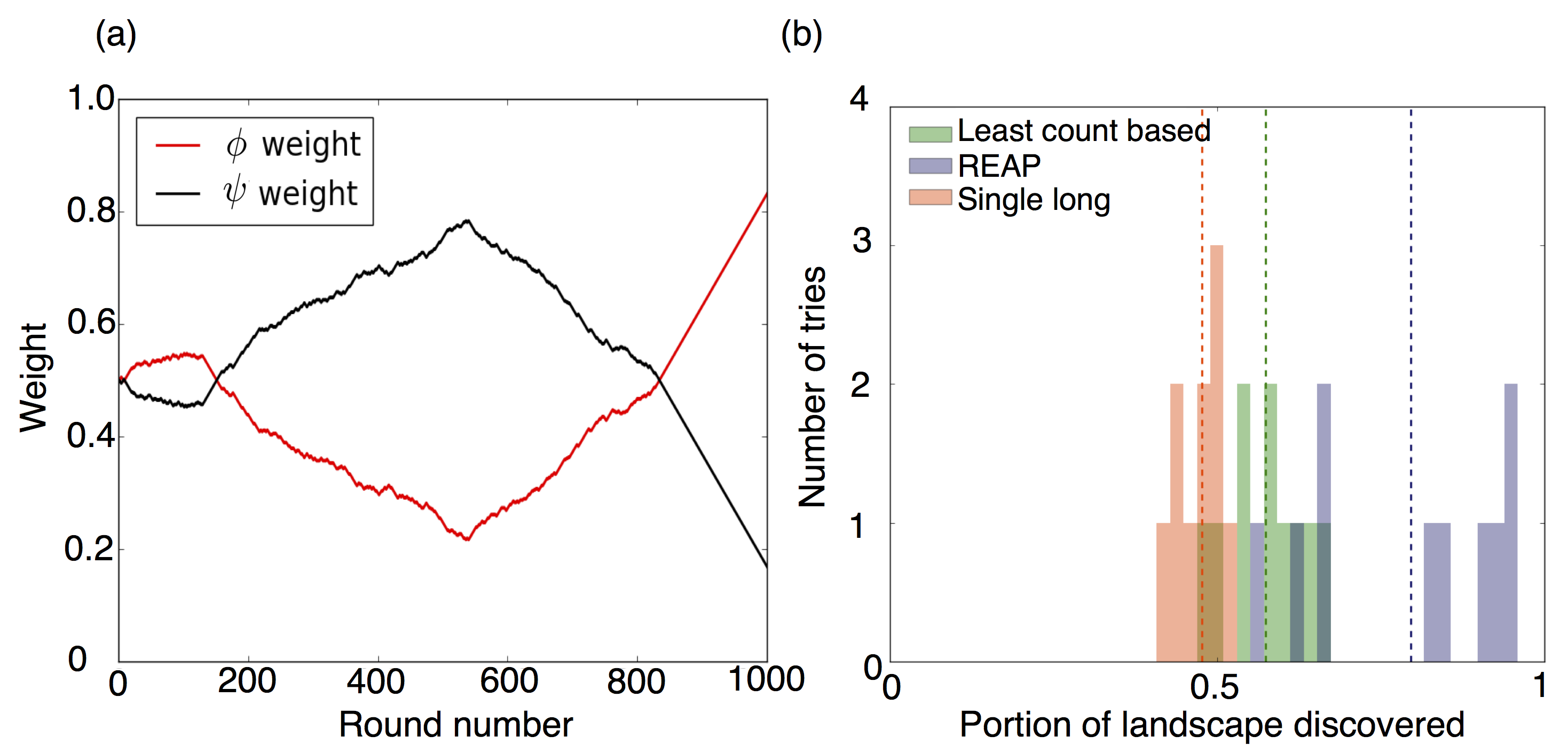}
\caption{(a) The weights corresponding to alanine dipeptide RCs ($\phi$ and $\psi$) change over each simulation round. As a result, these change in weights demonstrate how the algorithm identifies important direction of sampling as it explores the landscape (b) A plot showing the distribution of portion of landscape discovered over 10 repeated simulation trials. Dashed lines indicate the expected value for each distribution.}
\label{fig:ala-w}
\end{figure}

With regard to directionality, the algorithm initially starts to $\phi$ direction, but quickly learns that the RC that is best to sample along is along the $\psi$ coordinate (Fig. \ref{fig:ala-w} (a)). This change in weights effectively allowed the discovery of new regions along $\psi$.  At the second intersection shown in the plot in Fig. \ref{fig:ala-w} near round number 800 , the least-counts sampling built into REAP (step 5) allows for sampling of an entirely new region (with $\phi$ taking a value near ${\pi}/{4}$). The distribution in Fig. \ref{fig:ala-w} (b) illustrates that the REAP algorithm can be expected to outperform LC and SL upon inspecting the resulting expected values. It appears that two distributions arise from using REAP; one centers around 0.85 and the other around 0.6 near the LC distribution. This can be interpreted to indicate that it is unlikely that REAP will ever perform worse than LC. In addition, distributions of the individual angles, $\phi$ and $\psi$, were provided to demonstrate that REAP samples transitions better (SI Fig. 8 and 9). 

We considered the performance of REAP using two extra OPs, $\theta$, and $\zeta$. Much like the results with the L landscape (SI Fig. 4), sampling of 4 OPs is comparable with 2 OPs (SI Fig. 10). Moreover, distributions with respect to $\phi$ and $\psi$ remained alike (SI Fig. 11 and 12). These data indicate that REAP's performance does not diminish as extra OPs are added. 

\subsection{Src Kinase}
We furthermore demonstrated that the REAP algorithm's effectiveness in a protein system that has implications in cancer drug discovery. To provide some background, protein kinases are a famaily of enzymes that catalyze the transfer of phosphate group to serine, threonine, or tyrosine residues. \cite{shan2014,lin2013explaining,shukla2014activation} When the activity of this kinase becomes deregulated (perhaps due to a genetic mutation), it can cause uncontrolled cell proliferation leading to tumor development. Extensive analysis has been done previously using markov state models (MSMs) \cite{pande2010everything,shukla2015markov} to characterize the kinase dynamics \cite{shukla2014activation}. Furthermore, it has been shown that the conformational dynamics can be projected onto two OPs, forming an L-shaped landscape \cite{moffett2017molecular,shukla2014activation}. This system is therefore appropriate to apply REAP, as it is a natural extension of the aforementioned L model potential. Using the same MSM from this work\cite{shukla2014activation}, we generated trajectories using kinetic monte carlo (KMC) sampling. This sampling scheme uses the kinetics derived from the MSM to stochastically propogate the dynamics over time as opposed to integrating Newton's laws of motion. 

\begin{figure}[ht]
\centering
\includegraphics[width=\linewidth]{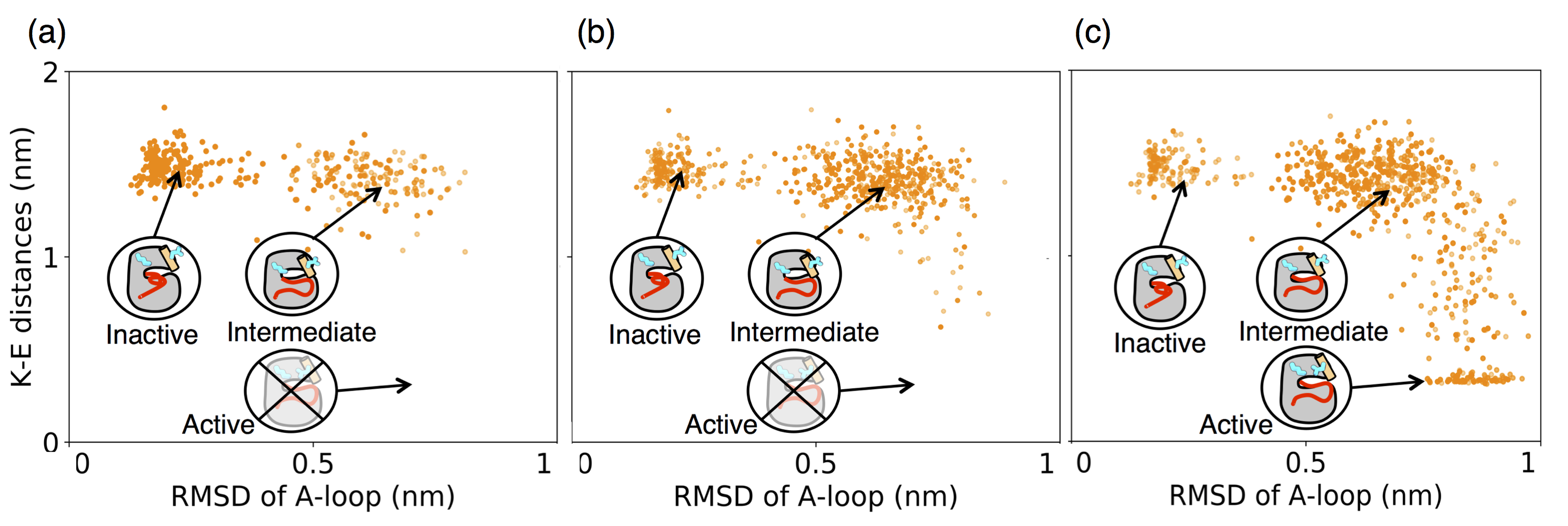}
\caption{(a) The simulation data of Src kinase using a single trajectory approach is plotted using two RCs: K-E distances and RMSD of the A-loop. Simulations started in the inactive conformation and ran for a total time of 15 $\mu$s. (b) Simulations data using least counts adaptive sampling. Sampling improves on (a), but fails to reach the active state (c) This plot shows how the REAP sampling algorithm outperforms that of the single trajectory approach in (a). The algorithm was able to facilitate sampling of the active kinase conformation.}
\label{fig:src2d}
\end{figure}

We used the python package MSMBuilder version 3.6\cite{harrigan2017msmbuilder} to carry out our stochastic simulations. Using KMC, we generated 15 $\mu$s of long continuous simulations and compared that to using the REAP algorithm. All of the simulations started from the same MSM state (the inactive kinase conformation), and we showed that the long simulation approach could not sample the active state within the given amount of time (Fig. \ref{fig:src2d} (a) and SI Fig. 13 for additional plots). The LC strategy (Fig. \ref{fig:src2d} (b) and SI Fig. 14) demonstrated improved sampling for the intermediate state, and the region between the active and intermediate. REAP, on the other hand, was able to discover an entire new area of the landscape corresponding to the active kinase conformation (Fig. \ref{fig:src2d} (c) and SI Fig. 15), a region that the other two methods failed to sample.

\begin{figure}[ht]
\centering
\includegraphics[width=\linewidth]{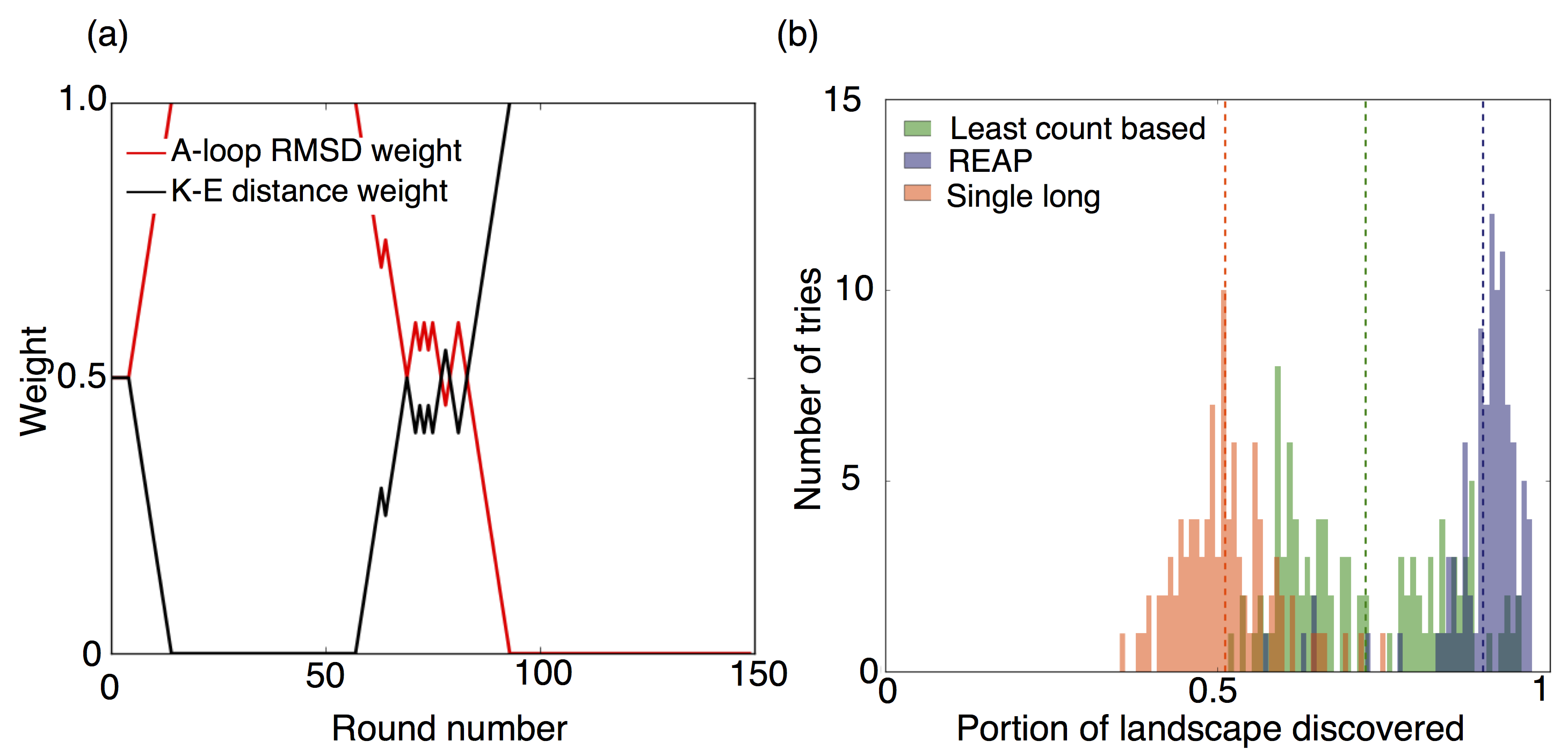}
\caption{(a) Weights for each RC signify the importance of each RC depending on the round number (or iteration of the algorithm) (b) A plot showing the frequency of portion of landscape discovered for REAP, LC, and SL simulations for 100 trials. Dashed lines represent the expected value of each distribution.}
\label{fig:src_weights_plts}
\end{figure}

A plot of the weight fluctuations are shown in Fig. \ref{fig:src_weights_plts} (a). It shows how the algorithm initially finds sampling along the A-loop RMSD more important than K-E distance (round 20-60). Afterwards, the weights fluctuate about 0.5 until the K-E distance then becomes relevant for sampling from 90 and onwards to reach the active conformation. The efficacy of the REAP algorithm is demonstrated in Fig. \ref{fig:src_weights_plts} (b) which shows a distribution of portion of landscape discovered for 100 repeated trials. The expected values are nearly equally spaced, with REAP's distribution aggregating around 0.9. The distribution for LC is split, centering at 0.6 and 0.8. On the other hand, SL is shifted to the left, concentrated at 0.5. Given this data, REAP can be expected to explore the landscape faster than SL and LC. 

Not only is REAP efficient at landscape exploration, but it also reaches the active state of Src kinase in less time (SI Fig. 16, 17, and 18). REAP and LC both perform better than SL, with about half the median value for the time to reach active. REAP is further shifted towards shorter times to reach the active state compared to LC, accompanied by a difference of about 5 $\mu$s (SI Fig. 19). Furthermore, REAP samples the active state and transition better than LC and SL. This is evident from, the distribution of K-E distances shown in SI Fig. 20.

We were further interested in determining the efficiency of our algorithm when extra non-functional order parameters (i.e. distances that do not change significantly over time) are considered. This was achieved by first providing 1 insignificant distance on the $\alpha$E helix (3 OPs total, SI Fig. 21), then, in a separate case, 10 additional distances situated on the $\alpha$E and $\alpha$F helix\cite{Roskoski2015} (12 OPs total, SI Fig. 22). 

The time to reach active using 3 and 12 OPs are comparable with using only 2 OPs by a median difference of 2 and 3 $\mu$s respectively (SI Fig. 23). Despite this difference, 3 and 12 OPs still maintain a faster median time than LC. We moreover plotted the time series of the average smallest K-E distance for each instance illustrating that introducing these non-productive OPs will still perform better than LC over time (compare SI Fig. 17 and 26). Similarly, distributions of the K-E distance for cases with multiple OPs still sample active state better than LC (compare SI Fig. 18 and 27).

When contrasting the distribution for the portion of landscape discovered using 3 and 12 OPs (SI Fig. 24), we found that the expectation values are still greater than that of the LC sampling with 2 OPs. Further, the distributions contain overlapping regions, with an expectation value difference of only $\sim 0.1$. These results, as well as the ones of alanine dipeptide and L-shaped landscape (SI Fig. 10 and 4 respectively), suggest that the introduction of additional OPs will not dramatically decrease the performance of the algorithm, and can still be expected to perform better than LC. With regard to how active state sampling is affected when introducing multiple OPs, our results indicate that these superfluous distances bear no affect (compare SI Fig. 20 (c) and Fig. 25). REAP successfully attributes low weights to these additional/irrelevant distances as their values remain low throughout the simulations (SI Fig. 28 and 29).

\par\noindent  The proposed algorithm, REAP, has been shown to efficiently sample landscapes in the case of both model potentials (L-shaped, circular), alanine dipeptide, and Src kinase. It achieves this by identifying which RCs maximizes a reward function that encourages exploratory behavior. This is mathematically represented as weights and we have demonstrated that the algorithm is able to determine which RCs are preferable while exploring the conformational landscape. In all systems that were studied, REAP consistently outperformed the traditional simulation approach and LC sampling when examining the distribution landscape discovered for the same simulation time.

\par\noindent Regarding algorithm improvement, it is possible to introduce  multiple structures at different positions along the landscape, essentially allowing the simulations to be explored from different starting points. This idea is motivated from the concept of ``multi-agent reinforcement learning'' \cite{watkins1992q, littman1994markov, hu1998multiagent} by which agents can either interact in a cooperative or competitive fashion.  One drawback of REAP is that the selection of initial RCs will most likely depend on structural or biophysical data. However, the algorithm allows for a use of a large number of RCs and the algorithm reduces the weight associated with the fast directions to zero within a few rounds of sampling. Another possible way of choosing RCs is using the approach outlined by Tajkhorshid and coworkers that involves estimating the work done by performing short pulling simulations along the RC directions associated with the conformational change process.\cite{li2015computational, moradi2013mechanistic} If none are available, then one possibility is to use evolutionary coupling (EC) pairs\cite{hopf2014sequence} as RCs (distance between residues that evolve together over time) as RCs for the given sampling policy. 

\par\noindent In our recent work\cite{shamsi2017enhanced, feng2018characterizing}, we have shown that using evolutionary coupling distances as a criteria for least-counts adaptive sampling can enhance the exploration of the landscape. Given that the REAP algorithm uses this count based sampling strategy, we expect that using ECs as RCs for REAP will not only sample the landscape faster, but also differentiate between ECs that are actually relevant for conformational dynamics from those that are only important for protein folding. The separation between these two types of EC is still an open scientific question to the community. 

\par\noindent We believe this algorithm will be particularly beneficial for those interested in building MSMs. This is because it uses the swarm of simulation approach, which essentially runs many small trajectories in parallel. MSMs are the preferred theoretical framework at the moment to merge these discontinuous simulations and accurately reproduce the same observables as traditional MD. Additionally, REAP has advantages other than building MSMs. The principle challenge with most biased MD methods is that the original Hamiltonian is altered to preferentially sample some subset of the high dimensional space of proteins. The result of this alteration will then modify the probability distributions of protein configurations, with the possibility of favoring states that are less likely in actual biological systems. Therefore, observations on the dynamics of these biased simulations may not be useful for predicting the detailed kinetic or thermodynamic mechanism of conformational change in proteins.

\begin{acknowledgement}
Authors acknowledge the Blue Waters sustained-petascale computing project, which is supported by the National Science Foundation (awards OCI-0725070 and ACI-1238993) and the state of Illinois. Z.S. was supported by the Widiger and 3M fellowships from the Department of Chemical \& Biomolecular Engineering at University of Illinois, Urbana-Champaign, USA. K.J.C. is supported by the Ford Foundation fellowship from the National Academies of Arts, Sciences and Medicine. We also thank Stanford Digital Repository for providing the Src kinase MD simulation data. 
\end{acknowledgement}

\clearpage
\newpage

\providecommand{\latin}[1]{#1}
\providecommand*\mcitethebibliography{\thebibliography}
\csname @ifundefined\endcsname{endmcitethebibliography}
  {\let\endmcitethebibliography\endthebibliography}{}


\end{document}